\documentclass[a4paper,11pt]{article}
\usepackage{pos-vaso}
\usepackage{url,units,xspace}
\usepackage{lineno}

\newcommand{\ifb}{\mbox{fb$^{-1}$}\xspace}

\newcommand{\gev}{\mbox{GeV}\xspace}
\newcommand{\tev}{\mbox{TeV}\xspace}

\newcommand{\gd}{\ensuremath{g_{\textrm D}}\xspace}
\newcommand{\half}{\nicefrac{1}{2}\xspace}

\newcommand{\none}{\ensuremath{\tilde{\chi}_1^0}\xspace}

\DeclareGraphicsExtensions{.pdf,.png,.jpg}
\graphicspath{{./}{./figs/}}

\title{Results and future plans of the MoEDAL experiment}

\author*[a]{Vasiliki A.\ Mitsou}
\author{for the MoEDAL Collaboration}

\affiliation[a]{Instituto de F\'isica Corpuscular (IFIC), CSIC -- Universitat de Val\`encia, \\ 
 C/ Catedr\'atico Jos\'e Beltr\'an 2, E-46980 Paterna (Valencia), Spain}

\emailAdd{vasiliki.mitsou@ific.uv.es}

\abstract{The unprecedented collision energy of the LHC has opened up a new discovery frontier. Unfortunately, signs of new physics have yet to be seen. The LHC's first dedicated search experiment, MoEDAL, started data taking for LHC Run~2. MoEDAL is designed to search highly ionising particle avatars of new physics using $pp$ and heavy-ion collisions at the LHC. The planned upgrade for MoEDAL at Run~3 --- the MAPP detector (MoEDAL Apparatus for Penetrating Particles) --- will extend MoEDAL's physics reach to include feebly interacting and long-lived messengers of physics beyond the Standard Model. This will allow us to explore a number of models of new physics, including dark-sector models, in a complementary way to that of conventional LHC collider detectors. This article focuses on recent results and plans for the LHC Run~3.}

\FullConference{%
  *** The European Physical Society Conference on High Energy Physics (EPS-HEP2021), ***\\
  *** 26-30 July 2021 ***\\
  *** Online conference, jointly organized by Universität Hamburg and the research center DESY ***
}

\begin{document}
\maketitle


\section{Introduction}\label{intro}

The MoEDAL (Monopole and Exotics Detector at the LHC)~\cite{Pinfold:2009oia} experiment at the Large Hadron Collider (LHC) is mainly dedicated to searches for manifestations of new physics through highly ionising particles in a manner complementary to ATLAS and CMS~\cite{DeRoeck:2011aa}. It is the  first dedicated \emph{search} LHC experiment to be approved with others following~\cite{Alimena:2019zri}. The principal motivation for MoEDAL is the quest for magnetic monopoles and dyons, as well as for any massive, stable or long-lived, slow-moving particle with single or multiple electric charge arising in various extensions of the Standard Model (SM)~\cite{Acharya:2014nyr}.  Emphasis is given here on recent MoEDAL results, based on the exposure of magnetic monopole trapping volumes to 13~\tev $pp$ collisions, on future prospects, including electric charges, and on the description and sensitivity of the planned MoEDAL Apparatus for Penetrating Particles (MAPP), designed to extend the LHC reach in the quest for dark matter~\cite{Mitsou:2013rwa}.

\section{The MoEDAL detector}\label{sc:detector}

The MoEDAL detector~\cite{Pinfold:2009oia,Mitsou:2020hmt} is deployed around the intersection region at Point~8 (IP8) of the LHC in the LHCb Vertex Locator cavern. It is a unique and, to a large extend, passive detector based on three different detection techniques. 

The main sub-detector system is made of a large array of CR39\textregistered,  Makrofol\textregistered\ and Lexan\textregistered\ nuclear track detector (NTD) stacks surrounding the IP8. The passage of a highly ionising particle through the plastic sheets is marked by an invisible damage zone along the trajectory. The damage zone is revealed as a cone-shaped etch-pit when the plastic detector is chemically etched in the INFN Bologna laboratory. Then the sheets of plastics are scanned looking for aligned etch pits in multiple sheets. 

A unique feature is the use of magnetic monopole trappers (MMTs) to capture electrically and magnetically charged particles. The aluminium absorbers of MMTs are subject to an analysis looking for magnetically charged particles at the ETH Zurich SQUID laboratory~\cite{DeRoeck:2012wua}. A search for decays of \emph{electrically} charged objects stopped in the MMTs can subsequently be carried out at an underground facility in the MoEDAL Apparatus for very Long Lived particles (MALL)~\cite{Pinfold:2019zwp,Pinfold:2019nqj}.

The only active MoEDAL sub-detector comprises an array of several TimePix pixel devices dedicated to the monitoring of cavern background sources. Its time-over-threshold mode allows a 3D mapping of the charge spreading effect in the silicon sensor volume, thus differentiating between different particles species from mixed radiation fields and measuring their energy deposition~\cite{MOEDAL:2021mrd}.

\section{Searches for monopoles and dyons in MoEDAL}\label{sc:lightsearch}

MoEDAL is designed to fully exploit the energy-loss mechanisms of magnetically charged particles~\cite{Dirac:1931kp,Dirac:1948um,tHooft:1974kcl,Polyakov:1974ek}  in order to optimise its potential to discover these messengers of new physics. Various theoretical scenarios foresee the production of magnetic charge at the LHC~\cite{Acharya:2014nyr,Mavromatos:2020gwk}: (light) 't Hooft-Polyakov monopoles~\cite{tHooft:1974kcl,Polyakov:1974ek,Vento:2013jua}, electroweak monopoles~\cite{Cho:1996qd,Bae:2002bm,Cho:2012bq,Cho:2016npz,Ellis:2016glu,Ellis:2017edi,Alexandre:2019iub}, global monopoles~\cite{Barriola:1989hx,Drukier:1981fq,Mazur:1990ak,Mavromatos:2017qeb,Mavromatos:2016mnj,Mavromatos:2018drr,Mavromatos:2018kcd} and monopolium~\cite{Dirac:1948um,Zeldovich:1978wj,Hill:1982iq,Dubrovich:2002gp,Epele:2012jn,Vento:2020vsq}, a monopole-pair bound state. Magnetic monopoles and dyons, possessing both magnetic and electric charge~\cite{Schwinger:1969ib}, are fascinating hypothetical particles. Even though there is no empirical evidence for their existence, strong theoretical reasons hint that they do exist and many theories, including grand unified theories and superstring theory, predicted them.

Up to now, the MoEDAL physics results on monopoles are based on the scanning of the MMTs, exposed to LHC Run~1 8-\tev data~\cite{MoEDAL:2016jlb} and to 13~\tev $pp$ collisions~\cite{Acharya:2016ukt,Acharya:2017cio,Acharya:2019vtb}. The SQUID analysis yielded no observed isolated magnetic charges, leading to upper limits on monopole production cross sections. This outcome led to lower mass exclusion bounds when considering two pair production processes: (a) a Drell-Yan-like (DY) process in photon $s$-channel intermediation, and (b) a photon-fusion $t$-channel diagram~\cite{Baines:2018ltl}. 

\begin{figure}[htb]
   \centering
   \begin{minipage}[b]{0.58\linewidth}
   \includegraphics[width=\linewidth]{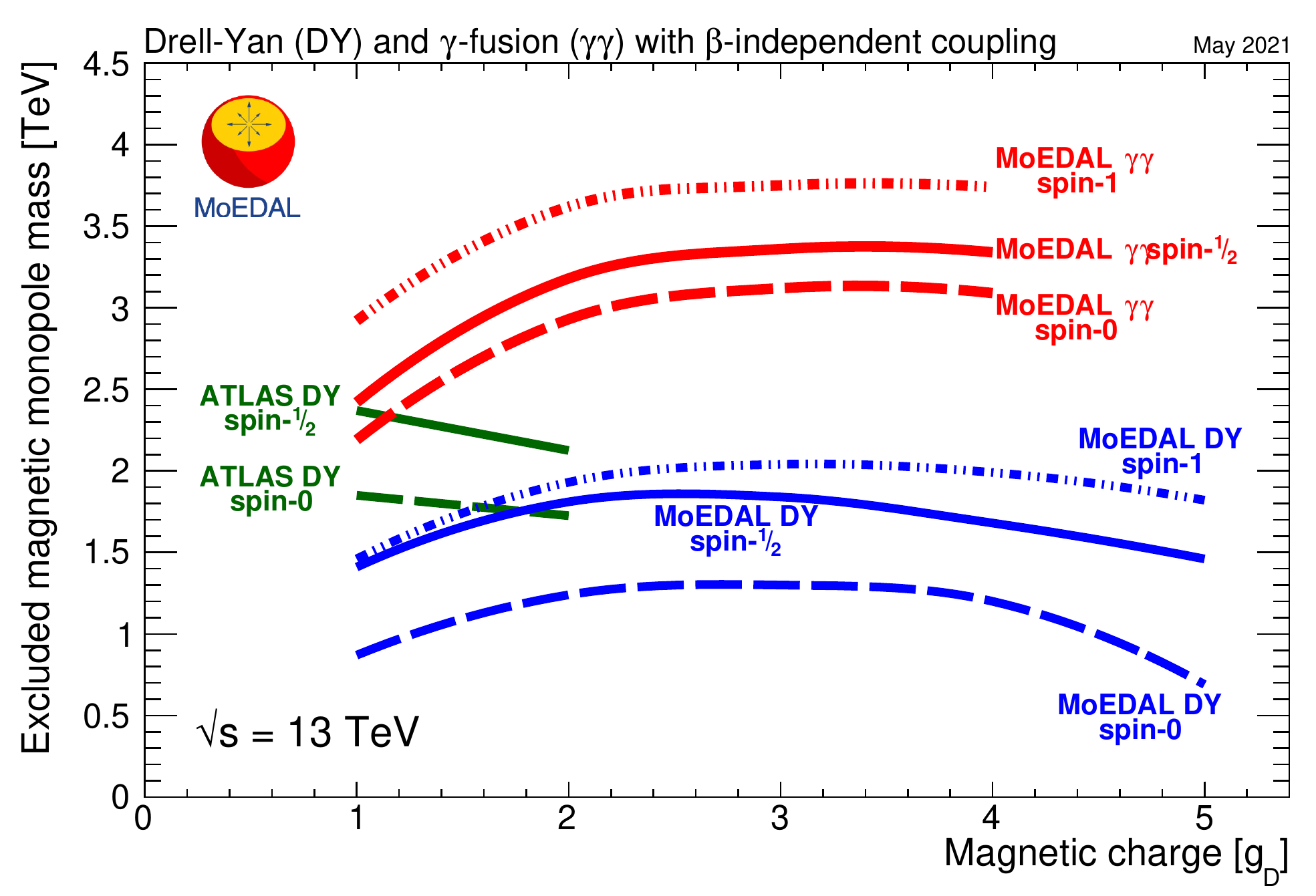}
   \end{minipage}\hspace{0.03\linewidth}
\begin{minipage}[b]{0.38\linewidth}
   \caption{Magnetic monopole mass limits from ATLAS~\cite{Aad:2019pfm} and MoEDAL searches~\cite{MoEDAL:2016jlb,Acharya:2019vtb} at $\sqrt{s} = 13~\tev$ as a function of magnetic charge for various spins, assuming  $\beta$-independent coupling and two pair-production mechanism: Drell-Yan and photon fusion. }
   \label{fg:moedal-dy-photon-fus-may2021}
   \end{minipage} 
\end{figure}

A comparison between the photon-fusion and the DY mass limits is presented in Figure~\ref{fg:moedal-dy-photon-fus-may2021} together with bounds set by ATLAS~\cite{Aad:2012qi,Aad:2015kta,Aad:2019pfm}.  The ATLAS bounds are better that the MoEDAL ones for $|g|=2\gd$ due to the higher luminosity delivered in ATLAS and the loss of acceptance in MoEDAL for small magnetic charges, while MoEDAL is the sole detector sensitive to high charges. The production cross section at the LHC energies for photon fusion is much higher than the DY, hence the more stringent limits of the former. 

\begin{figure}[ht]
\centering
\begin{minipage}[b]{0.5\linewidth}
  \includegraphics[width=\textwidth]{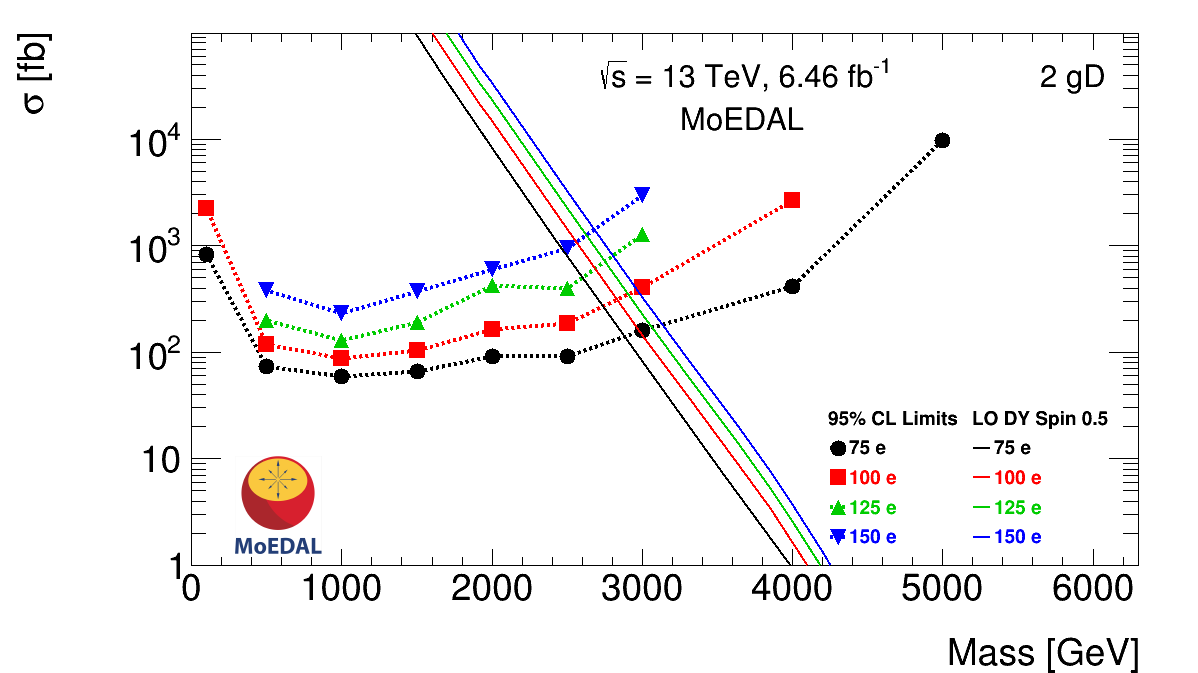}
  \caption{95\% CL cross-section upper limits for Drell-Yan dyon pair production in 13~\tev $pp$ collisions as a function of mass for spin-\half dyons with magnetic charge 2\gd. The solid lines are cross-section calculations at leading order~\cite{MoEDAL:2020pyb}.} \label{fg:dyon-spinhalf-2gd}
\end{minipage}\hspace{0.03\linewidth}
\begin{minipage}[b]{0.46\linewidth}
  \includegraphics[width=0.95\textwidth]{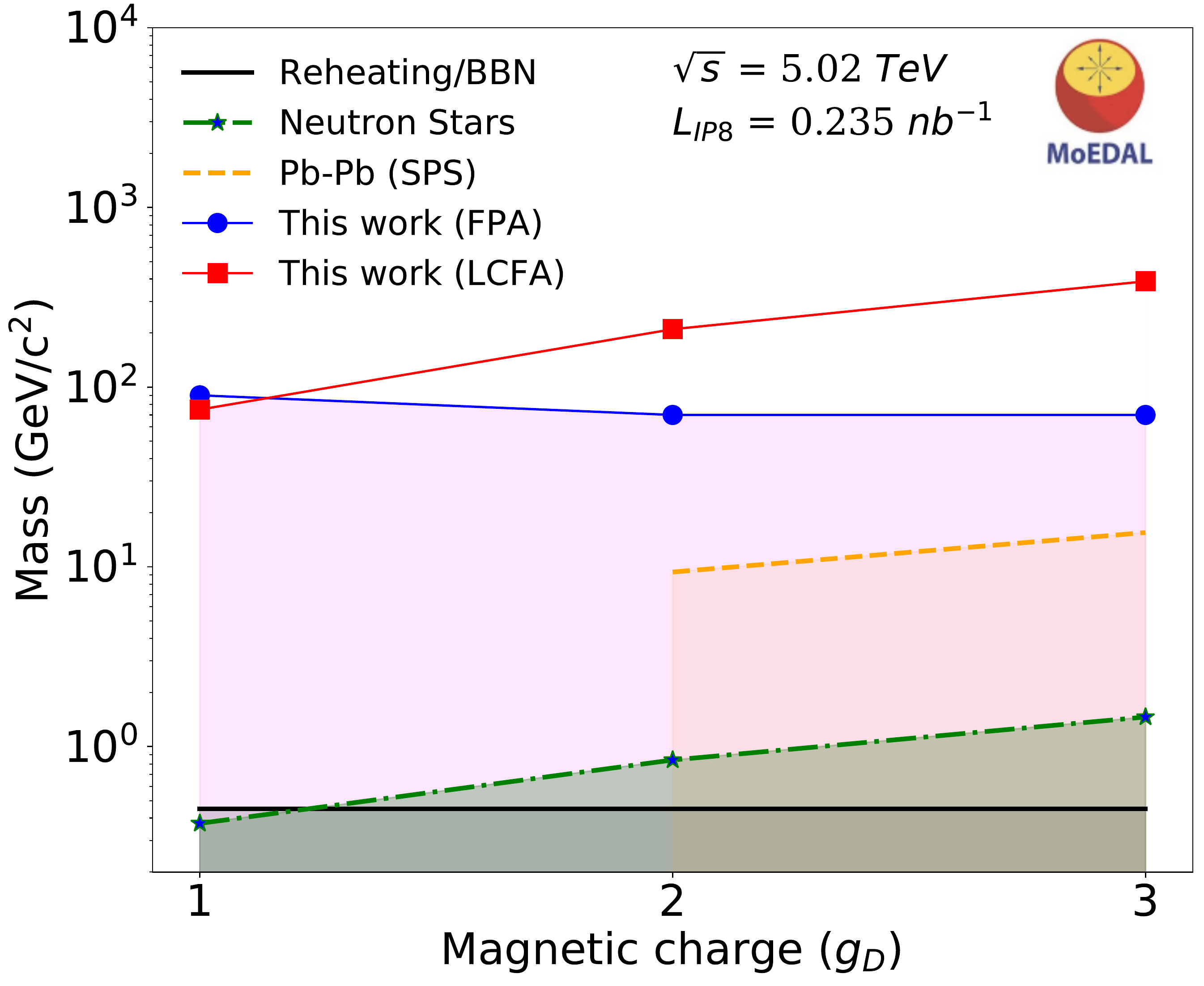}
  \caption{95\% CL monopole exclusion region (violet) obtained using two different calculations of Schwinger production mechanism for Pb-Pb collisions~\cite{Acharya:2021ckc}.} \label{fg:massbounds_schwingermonopoles}
\end{minipage}   
\end{figure}

MoEDAL performed recently the first dedicated dyon search in a collider experiment by means of MMT scanning; an example of upper cross section limits is shown in Figure~\ref{fg:dyon-spinhalf-2gd}. Mass limits in the range $750 - 1910~\gev$ were set using a benchmark DY production model for dyons with magnetic charge up to 5\gd, for electric charge from $1e$ to $200e$, and for spins 0, \half and 1~\cite{MoEDAL:2020pyb}. 

In both production processes, the monopole pair couples to the photon via a coupling that depends on $\gd$ and hence it is ${\cal O}(10)$. This large monopole-photon coupling invalidates any perturbative treatment of the cross-section calculation and hence any result based on it is \emph{only indicative} and used merely to facilitate comparisons between experiments. On the contrary, the upper bounds placed on production cross sections are solid and can be relied upon.

This situation may be resolved if thermal Schwinger production of monopoles in heavy-ion collisions is considered~\cite{Schwinger:1951nm}. This mechanism becomes effective in the presence of strong magnetic fields and calculations rely on semiclassical techniques rather than perturbation theory,  therefore it overcomes these limitations~\cite{Gould:2017zwi,Gould:2017fve,Gould:2018efv,Gould:2019myj,Ho:2019ads,Gould:2021bre}. Pb-Pb heavy-ion collisions at the LHC produce the strongest known magnetic fields in the current Universe, and the first search for such production was conducted by MoEDAL during the 5.02 TeV/nucleon heavy-ion run, during which the MMTs were exposed to 0.235~nb$^{-1}$ of Pb-Pb collisions and analysed later with a SQUID. Monopoles with Dirac charges $1\gd \leq g \leq 3\gd$ and masses up to 75~\gev were excluded, as seen in Figure~\ref{fg:massbounds_schwingermonopoles}. This provides the first lower mass limit for finite-size monopoles from a collider search~\cite{Acharya:2021ckc}.

\section{Beyond magnetic monopoles: electrically charged particles}\label{sc:heco}

The MoEDAL detector is also designed to search for any massive, long-lived, slow-moving particles with single or multiple electric charges arising in many scenarios of physics beyond the Standard Model. Supersymmetric (SUSY) long-lived particles~\cite{Mavromatos:2016ykh,Mitsou:2020hmt,Sakurai:2019bac,Felea:2020cvf}, quirks, strangelets, Q-balls, and many others fall into this category~\cite{Acharya:2014nyr}. A generic search for high-electric-charge objects (HECOs) is currently underway using for the first time NTD data~\cite{moedal-hecos}. 

\begin{figure}[ht]
\centering
\begin{minipage}[b]{0.5\linewidth}
  \includegraphics[width=\textwidth]{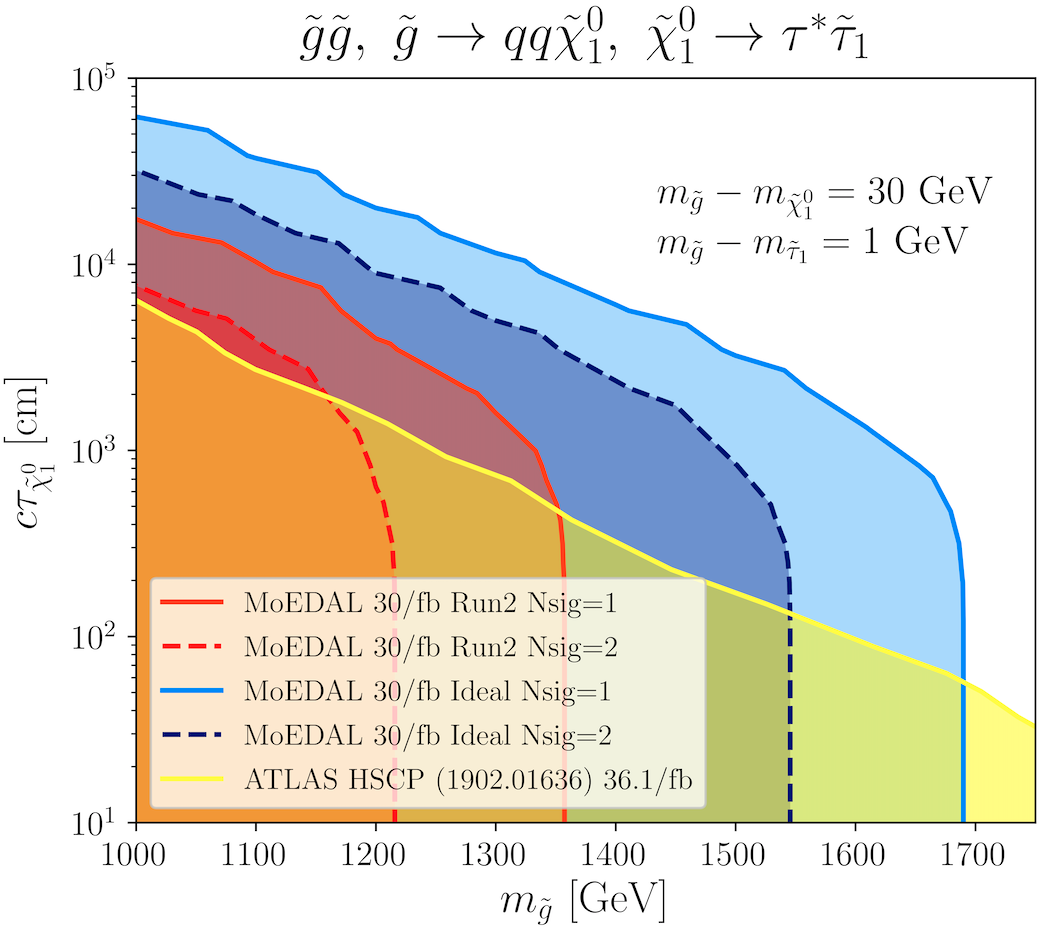}
  \caption{MoEDAL sensitivity in sleptons produced in gluino pair production, for two different NTD geometries. The yellow shaded region is excluded by ATLAS~\cite{Aaboud:2019trc}. From~\cite{Felea:2020cvf}.} \label{fg:comb_signal_SUSY}
\end{minipage}\hspace{0.03\linewidth}
\begin{minipage}[b]{0.46\linewidth}
  \includegraphics[width=\textwidth]{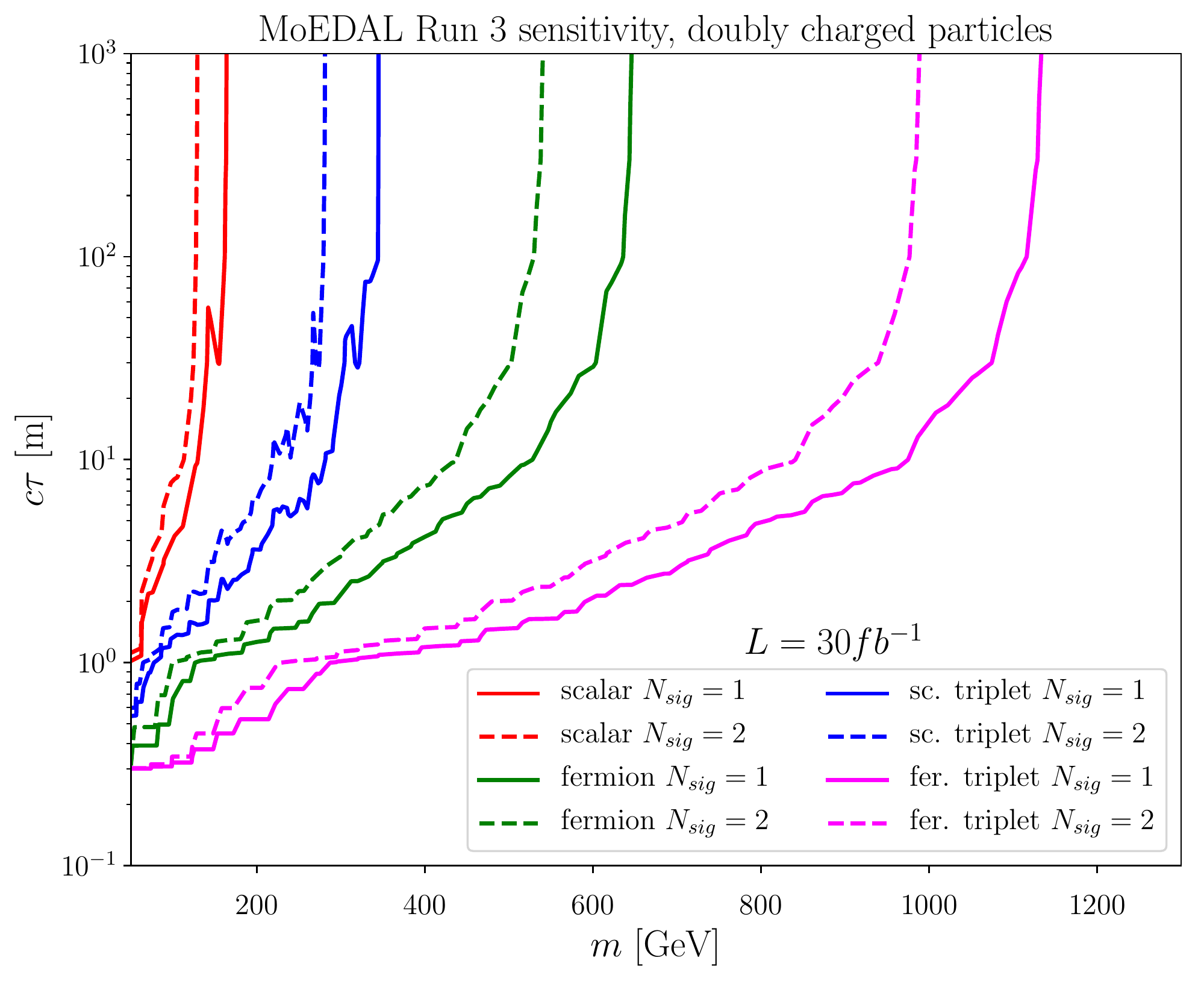}
  \caption{The expected sensitivities of MoEDAL for various long-lived doubly charged particle species, assuming 30~\ifb of integrated luminosity~\cite{Acharya:2020uwc}.} \label{fg:comb_signal_2Q}
\end{minipage}   
\end{figure}

A feasibility study on the detection of massive metastable supersymmetric partners showed that MoEDAL is mostly sensitive to slow-moving particles unlike ATLAS/CMS suitability for faster ones, yet the less integrated luminosity it receives at IP8 remains a limiting factor for simple scenarios. Direct production of heavy fermions with large cross section (thus via strong interactions) is the most favourable scenario for MoEDAL. More complex topologies appear to be promising for detecting long-lived sleptons in phenomenologically realistic models, as shown in Figure~\ref{fg:comb_signal_SUSY}, where MoEDAL could cover parameter space less accessible by CMS~\cite{Sakurai:2019bac} and ATLAS~\cite{Felea:2020cvf} . Even for SUSY models observable by both ATLAS/CMS and MoEDAL, the added value of MoEDAL would remain, since it provides a coverage with a completely different  detector and analysis technique, thus with uncorrelated systematic uncertainties.

The prospects for detecting particles of higher electric charges in MoEDAL are also very promising. Doubly charged scalars and fermions  are suggested by Type-II ($H^{\pm\pm}$) and Type-III seesaw models of neutrino masses, respectively. In addition to the Type-II seesaw model, several other BSM scenarios, namely the Left-Right model, the Georgi-Machacek model, the 3-3-1 model and the little-Higgs model also predict doubly-charged scalars. The expected MoEDAL sensitivity in terms of lifetime and mass is shown in Figure~\ref{fg:comb_signal_2Q}. Even better discovery reach is anticipated for charges of $2e$, $3e$ and $4e$, proposed in radiative neutrino mass models, which often add a discrete symmetry to the SM gauge group.~\cite{R:2020odv} For such models, at least one signal event at the MoEDAL NTDs is expected for up to masses of 290, 610 and 960~\gev for scalars $S^{\pm2}$, $S^{\pm3}$ and $S^{\pm4}$ in Run~3~\cite{Hirsch:2021wge}.

\section{MoEDAL Apparatus for Penetrating Particles}\label{sc:mapp} 
 
MoEDAL is proposing to deploy MAPP in a gallery near IP8 shielded by an overburden of approximately 100~m of limestone from cosmic rays~\cite{Pinfold:2019zwp}. It is envisaged that the first-stage detector, MAPP-1 will be installed during Long Shutdown~2 to take data in LHC Run~3. The purpose of the innermost detector, the MAPP-mQP, is to search for particles with fractional charge as small as $0.001e$, the so called millicharged particles (mCP), using plastic scintillation bars. A prototype of the mQP detector (10\% of the original system) is already in place since 2017 and the data analysis is progressing. The Phase-1 MAPP-mQP is going to be deployed in the UA82 gallery in a distance $100~{\rm m}$ from IP8.

\begin{figure}[ht]
\centering
\begin{minipage}[b]{0.5\linewidth}
  \includegraphics[width=\textwidth]{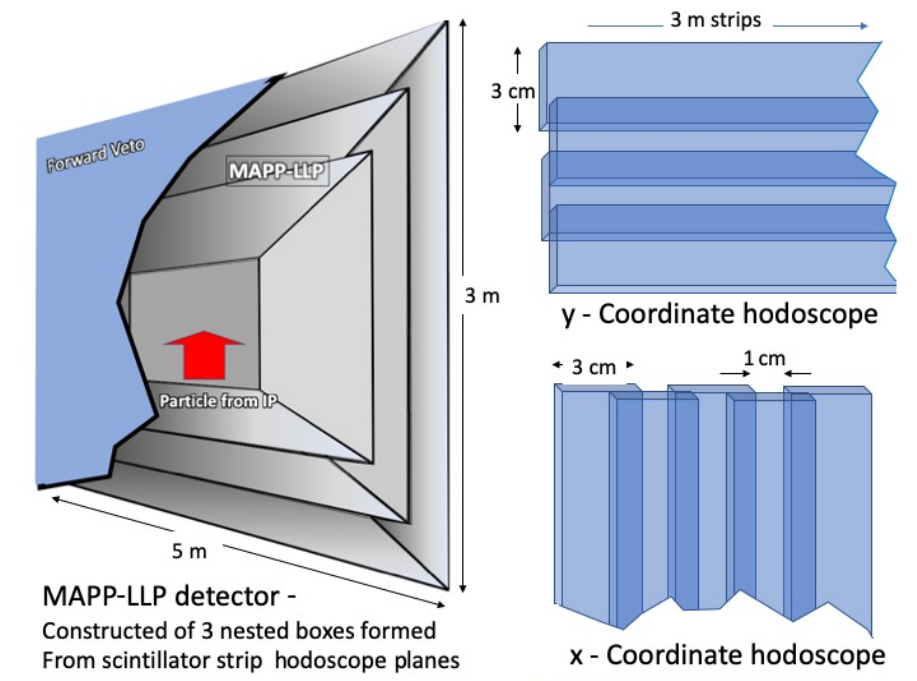}
  \caption{A schematic view of the MAPP-LLP subdetector.} \label{fg:mapp-llp}
\end{minipage}\hspace{0.03\linewidth}
\begin{minipage}[b]{0.46\linewidth}
  \includegraphics[width=\textwidth]{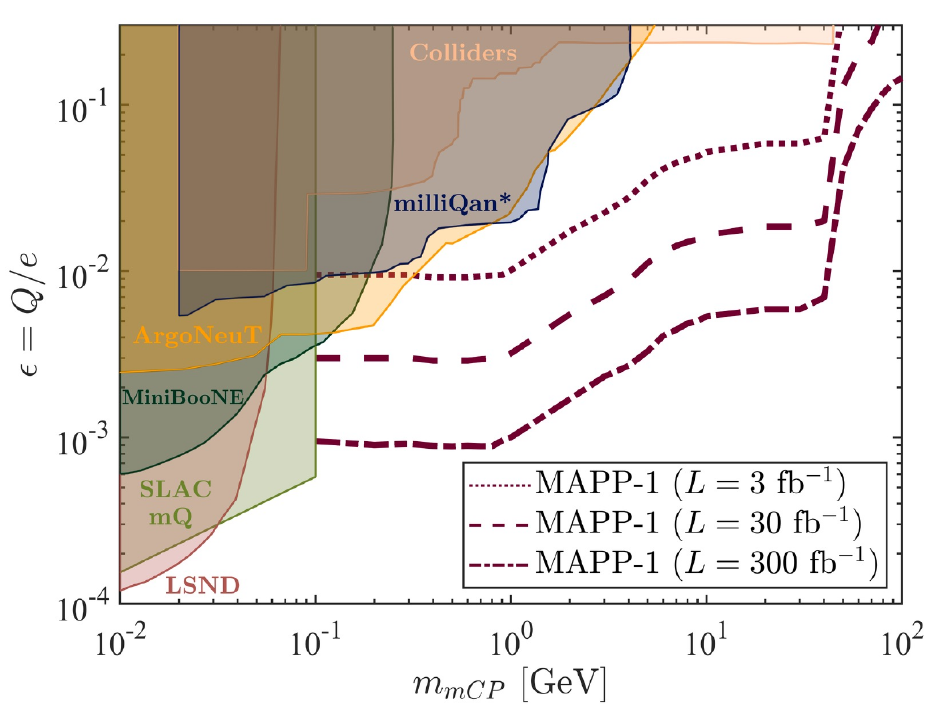}
  \caption{MAPP-mQP sensitivity to mCPs in a dark photon model for various integrated luminosity assumptions~\cite{Staelens:2021}.} \label{fg:dark-photon-mcp}
\end{minipage}   
\end{figure}

Another part of the detector, the MAPP-LLP, is deployed as three nested boxes of scintillator hodoscope detectors, in a `Russian doll' configuration, following as far as possible the contours of the cavern as depicted in Figure~\ref{fg:mapp-llp}. It is designed to be sensitive to long-lived neutral particles from new physics scenarios via their interaction or decay in flight in a decay zone of size approximately $5~{\rm m} \text{ (wide)} \times 10~{\rm m} \text{ (deep)} \times 3~{\rm m} \text{ (high)}$. The MAPP detector can be deployed in a number of positions in the forward direction, at at distance of $\mathcal{O}(100~{\rm m})$ from IP8. An upgrade plan for the MAPP-1 detector is envisaged for HL-LHC, called MAPP-2. 

mCPs arise when a new $U(1)$ is introduced to a dark sector with a massless dark photon $A'$, coupled to the SM photon field, and a massive dark fermion $\psi$ with charge much less than that of an electron as a result of kinetic mixing. MAPP-mQP is expected to significantly extend the reach obtained by milliQan demonstrator~\cite{Ball:2020dnx}, as shown in Figure~\ref{fg:dark-photon-mcp}. Moreover, it has been shown that MAPP-mQP can detect a heavy neutrino with a large enough electric dipole moment considered to be a member of a fourth generation lepton doublet~\cite{Frank:2019pgk}.
 
The MoEDAL-LLP detector should be sensitive to portal interactions that connect a hidden (dark) sector and the visible sector of the SM, leading to dark photons, dark Higgs bosons, heavy neutral leptons and axions. Scenarios beyond-the-SM that introduce a dark sector in addition to the visible SM sector are required to explain a number of observed phenomena in particle physics, astrophysics and cosmology such as the non-zero neutrino masses and oscillations, the dark matter, baryon asymmetry of the Universe, the cosmological inflation. In particular, dark Higgs bosons interact with the SM through a kinetic mixing term, thus probing one of the few possible renormalisable interactions with a hidden sector, the Higgs portal quartic scalar interaction. Such scenarios are accessible to MAPP and other future experiments~\cite{Mitsou:2020okk,Mitsou:2021tti}. Regarding their cosmological implications, dark Higgs bosons may mediate interactions with hidden dark matter that has the correct thermal relic density or resolves small-scale-structure discrepancies. Indeed, such a model with a dark Higgs inflaton strongly favoured by cosmological Planck+BK15 data is expected to leave imprints on LHC experiments, MAPP-LLP included~\cite{Popa:2021fgy}.

In the fermion portal, right-handed long-lived heavy neutrinos $N_i$ can be pair produced in the decay of an additional $Z'$ boson in the gauged $B-L$ model, which also contains a singlet scalar field that spontaneously breaks the extra $U(1)_{B-L}$ gauge symmetry. In this model, MAPP-2 will fill the gap left by other LHC experiments~\cite{Deppisch:2019kvs}. Sterile neutrinos may be long-lived in neutrino-extended SM Effective Field Theories, $\nu$SMEFT. Intermediate-mass can be produced in leptonic and semi-leptonic decays of charmed and bottomed mesons, decaying to leptons via neutral and charged weak currents, thus becoming detectable in MAPP-LLP~\cite{DeVries:2020jbs}. $R$-parity violating supersymmetry also predicts LLPs, such as light long-lived neutralinos \none decaying via $\lambda'_{ijk}$ couplings to charged particles. Benchmark scenarios related to either charm or bottom mesons decaying into \none have been considered, in similar fashion as in sterile neutrinos, showing that the MAPP can cover various \none lifetimes~\cite{Dreiner:2020qbi}.
 
\section{Summary and outlook}\label{sc:summary}

MoEDAL, the first dedicated \emph{search} LHC experiment, is extending considerably the LHC reach in the search for (meta)stable highly ionising particles. The latter are predicted in a variety of theoretical models and include magnetic monopoles, SUSY long-lived spartners, D-matter~\cite{Shiu:2003ta,Ellis:1998nz,Ellis:2008gg,Ellis:2004ay,Ellis:2005ib,Mavromatos:2010jt,Mavromatos:2010nk}, quirks, strangelets, Q-balls, etc~\cite{Acharya:2014nyr,Mavromatos:2016ykh}. MoEDAL is optimised to probe precisely all such long-lived states, unlike the other LHC experiments~\cite{DeRoeck:2011aa}, by combining different detector technologies: plastic nuclear track detectors, trapping volumes and pixel sensors. It provides the best limits in high magnetic charges.

Latest highlights in MoEDAL results include the first search for dyons at the LHC and the first search for monopoles produced via the Schwinger mechanism. The first search with NTDs, looking for HECOs~\cite{moedal-hecos}, and the analysis for trapped monopoles in the Run~1 CMS beam pipe is expected to be announced soon. MoEDAL is also sensitive to single and multiple electric charges predicted in supersymmetric models and scenarios explaining neutrino masses.

\acknowledgments
This work was supported in part by the Generalitat Valenciana via a special grant for MoEDAL and via the Project PROMETEO-II/2017/033, and by the Spanish MICIU / AEI and the European Union / FEDER via the grant PGC2018-094856-B-I00.

\bibliographystyle{JHEP-notitle}    
\small
\bibliography{skeleton}

\end{document}